\documentclass[a4paper,prd,noshowpacs,noshowkeys,notitlepage]{revtex4}
\usepackage[latin1]{inputenc}
\usepackage[T1]{fontenc}
\usepackage{ae,aecompl} 
\usepackage{amsmath,amssymb}
\usepackage{hyperref}

\providecommand{\la}[1]{\lambda_{#1}}

\providecommand{\dV}[1]{{\frac{\partial V}{\partial #1} }}
\providecommand{\dd}[2]{{\frac{\partial #1}{\partial #2} }}
\providecommand{\bx}{{\mathbf{x}}}

\providecommand{\bbB}{{\mathbb{B}}}
\providecommand{\tB}{{\tilde{\mathbb{B}}}}
\providecommand{\ums}[1]{\ml{\tfrac{1}{#1}}}
\providecommand{\eps}{{\epsilon}}
\providecommand{\tchi}{{\tilde{\chi}}}
\DeclareMathOperator{\rank}{rank}
\providecommand{\ones}[2]{{A_{#1\times #2}}}
\providecommand{\aver}[1]{\langle #1 \rangle}
\providecommand{\tp}{{\mss{\mathsf{T}}}}
\providecommand{\mss}[1]{\mbox{\scriptsize $#1$}}
\providecommand{\ml}[1]{\mbox{\large $#1$}}
\DeclareMathOperator{\im}{\mathrm{Im}} 
\providecommand{\mtrx}[1]{\begin{pmatrix} #1 \end{pmatrix}}
\DeclareMathOperator{\diag}{\mathrm{diag}} 

\providecommand{\eq}[1]{\begin{equation} #1 \end{equation}}

\providecommand{\eqali}[1]{\begin{equation}\begin{aligned} #1
    \end{aligned}\end{equation}}
\providecommand{\subeqali}[2]{\begin{subequations}\label{#1}\begin{align} #2
    \end{align}\end{subequations}}
\begin{document}
\title{
Absence of Spontaneous CP violation in Multi-Higgs Doublet Extension of MSSM
}
\author{R.~N.~Mohapatra}
\email{rmohapat@umd.edu}
\affiliation{Maryland Center for Fundamental Physics, Department of Physics, University of Maryland, College Park, MD 20742, USA
}
\author{C.~C.~Nishi}
\email{celso.nishi@ufabc.edu.br}
\affiliation{
Universidade Federal do ABC - UFABC,\\
Rua Santa Adélia, 166, 09.210-170, Santo André, SP, Brazil
}

\date{\today}
\begin{abstract}
 We show that in the multi-Higgs extension of minimal supersymmetric standard model,
which has N-pairs of Higgs doublets (called NHMSSM), it is impossible to break CP
spontaneously, if we do not allow for fine-tuning relations between parameters.
The result holds true even in the presence of spontaneous R-parity
breaking.
\end{abstract}
\maketitle

\section{Introduction}
\label{sec:intro}

Evidence for the existence of fundamental forces that distinguish between matter
and antimatter (CP violating forces) is abundant, both in the laboratory and in
cosmology.
In the laboratory, there are the celebrated discoveries of CP violation in K- and
B-systems whereas in cosmology, CP violation is an essential
ingredient in our understanding of a fundamental mystery of the Universe, i.e., the
asymmetry between its matter and antimatter content.
While this information has been accumulating for over half a century,
an understanding of the origin and nature of these forces responsible for them has
eluded physicists. What is  known is that in the standard model, CP violation can be
parameterized
in terms of a single phase in the Cabibbo-Kobayashi-Maskawa quark rotation matrix.
However, we do not know where the phase comes from. For instance it is not known whether it resides as an intrinsic phase in
the interactions of Higgs bosons responsible for particle masses (Yukawa interactions)  or it arises dynamically in the ground state
of the theory, even though all interactions in the model are CP symmetric prior to
symmetry breaking. 
The latter class of models go by the name of ``spontaneous CP violation"
(SCPV)\cite{tdlee} and is the subject of this article.

The fact that spontaneous CP violation can occur in multi-Higgs extensions of the standard model
is very well known\cite{books}. Typically, it is highly dependent on the structure of
the Higgs potential. For instance in the Standard Model (SM)
which has only one Higgs doublet, there is no physical phase in its vacuum
expectation value (VEV)  and hence no spontaneous CP violation. With two Higgs
doublets\,\cite{tdlee}, the ground state can break CP only if there are both
quartic $(\phi^\dagger_1\phi_2)^2$ as well as quadratic terms $\phi^\dagger_1\phi_2$
present in the potential. When either of them is absent, the ground state
corresponds to real VEVs of the Higgs doublets and there is no spontaneous CP
breaking. In three Higgs extensions of theSM, even if the quadratic terms are
absent, there can be spontaneous CP breaking\cite{weinberg,branco} --- manifested
through both the exchange of charged\,\cite{weinberg} or
neutral\,\cite{deshpande.77} Higgs bosons.
Since there is a widespread belief that the new scale physics includes
supersymmetry, it is of interest to study CP breaking in MSSM and its multi-Higgs
extensions.

It is well known that in supersymmetric theories, holomorphy of
the superpotential, which is a requirement for the theory to be
supersymmetric, considerably restricts the terms in the Higgs
potential.  It is therefore important to study whether in such
theories, spontaneous CP violation can occur.  Such studies have
indeed been carried out in simple extensions of MSSM and it has
been shown that if we do not allow fine-tuning relations between the
parameters of the superpotential, it is impossible to have
spontaneous CP violation with only one additional singlet added to
MSSM\cite{romao.86} or with four and six Higgs doublet superfields
(two or three pairs of $H_u,H_d$) extensions of
MSSM\,\cite{rasin.98}. The four-Higgs case remains incapable of
spontaneously breaking CP even if one singlet is added provided
that there are no dimensionful parameters in the superpotential
involving the singlet\,\cite{rasin.98}; otherwise SCPV is in fact
possible\,\cite{pomarol.92}.

The proof in the four- and six-Higgs case given in \cite{rasin.98} uses geometric
constructions.
In this paper, we provide an alternative algebraic proof of the same result for the
four Higgs doublet extension of MSSM and further show that our technique has the
advantage that it can be  extended to the N-Higgs extension for arbitrary
$N$, with the result that there is no spontaneous CP breaking even in this case. We
then show that this result remains true even in the presence of spontaneous
R-parity breaking, where the sneutrino fields acquire nonzero VEVs\cite{aulakh}.

This paper is organized as follows: in Sec. \ref{sec:conditions}, we discuss the
general strategy for treating this question. In Sec.\,\ref{sec:susy:gen}, we apply
this to the general N-HMSSM, which has N/2 pairs of $(H_u,H_d)$ fields,
and to the general N-HMSSM with spontaneous R-parity breaking;
in particular, the 4-HMSSM is treated in more detail in Sec.\,\ref{sec:4HMSSM} as
an illustration of the method.
We present our conclusions in Sec.\,\ref{sec:conclusions}.

\section{General multi-Higgs models and conditions for spontaneous CP violation}
\label{sec:conditions}

We consider extensions of standard model with $N$ Higgs doublets $\phi_a$ with
$a=1,\ldots,N$, denoted here as N-Higgs-doublet-models (NHDMs).
In general we can assume that only the neutral members of the Higgs doublets acquire
VEVs, which we parameterize as:
\eq{
\aver{\phi_a}=(0, v_a e^{i\delta_a})^\tp\,.
}
We can fix one of the phases $\delta_1=0$ from hypercharge invariance.
Then, there will be SCPV (in the real basis) if $\delta_a\neq 0,\pi$ for some $a\ge
2$ and nonzero $v_a$. Such violation breaks the canonical CP transformation
\eq{
\label{def:CCP}
\phi_a(t,\bx)\stackrel{\rm CCP}{\to } \phi_a^*(t,-\bx)\,.
}
For some cases, there might be other inequivalent CP transformations that could
remain as a symmetry (multiple CP symmetries). We do not treat this case here.

We are interested to study the minima of the potential involving the $N$ Higgs
doublets and their CP properties.
To build a SCPV model, it is necessary to begin with a CP invariant potential
before SSB.
There is a basis where CP invariance implies that all parameters of the potential are real when written in
terms of $\phi_a^\dag\phi_b$.
This is the \textit{real basis}.
Let us suppose that a CP invariant NHDM potential in this basis has the form
\eq{
\label{V=V1+W}
V=V_1(|\phi_a|^2)+W(\phi_a^\dag\phi_b)\,,\quad
|\phi_a|^2\equiv \phi_a^\dag\phi_a\,.
}
Hence, we are avoiding terms such as $|\phi_1|^2\phi_1^\dag\phi_2$.
We also assume we are seeking neutral minima and terms such as
$|\phi^\dag_1\phi_2|^2$ are equivalent to $|\phi_1|^2|\phi_2|^2$, hence, we include
them in $V_1$.

This form of the potential covers a large class of theories such as
the Weinberg model\,\cite{weinberg,branco} for SCPV and supersymmetric models where
the terms $\phi_a^\dag\phi_b$, $a\neq b$, are only contained in the quadratic
parts\,\cite{rasin.98}.
This can be also adapted for potentials, without trilinear terms, made up of the
neutral components of different multiplets.

Let us derive the extremum equations for \eqref{V=V1+W},
\eq{
\label{dV:V1+W}
\dV{\phi_{ak}^*}=\phi_{ak}\dd{V_1}{K_{aa}}+\sum_{b\neq
a}\phi_{bk}\dd{W}{K_{ba}}=0\,.
}
We have used the shorthand
\eq{
K_{ba}\equiv\phi_a^\dag\phi_b\,.
}

If we contract \eqref{dV:V1+W} with $\phi_{ak}^*$, we obtain
\eq{
\label{dV:K}
\sum_{k=1,2}\phi_{ak}^*\dV{\phi_{ak}^*}
=K_{aa}\dd{V_1}{K_{aa}} +\sum_{b\neq a}K_{ba}\dd{W}{K_{ba}}
=0\,.
}
The imaginary part of \eqref{dV:K} yields
\eq{
\label{dV:K:im}
\sum_{b\neq a}\im\Big(K_{ba}\dd{W}{K_{ba}}\Big)=0\,.
}
Spontaneous CP violation (SCPV) will be possible only if a solution for
\eqref{dV:K:im} is nontrivial, i.e.,
\eq{
\im(K_{ab})\neq 0 ~~\text{for some $a\neq b$},
}
without the need for any fine-tuning of parameters.

To comply with a successful EWSB, it is necessary that
\eq{
\sum_{a}K_{aa}\approx\frac{246\text{GeV}}{\sqrt{2}} \,,
}
which requires at least one of $K_{aa}$ non-null.
Let us choose, without loss of generality, $K_{11}\neq 0$.

In the following, we will restrict the part of the potential that depends on the
relative phases of the VEVs, i.e., $W$, to be either quadratic or quartic in the
doublet fields. The quadratic case includes all general terms but the quartic case
excludes further terms such as $\phi_1^\dag\phi_2\phi^\dag_1\phi_3$.
To specify further the class of NHDM potentials we are considering here, we use the
following parametrization for the potential:
\eq{
\label{V:2|4}
V=\sum_a\mu_a|\phi_a|^2+\ums{2}\sum_{a,b}c_{ab}|\phi_a|^2|\phi_b|^2+W\,.
}
The parameters (matrix) $c_{ab}$ are real and symmetric.

\subsection{Quadratic $W$}
\label{sec:W:2}

Let us choose
\eq{
\label{W:susy}
W=\sum_{a< b}\la{ab}\phi^\dag_a\phi_b+h.c.=\sum_{a,b}\la{ab}K_{ba}\,.
}
The parameters $\la{ab}$ (matrix) are real and symmetric, $\la{ba}=\la{ab}$, but we
adopt $\la{aa}=0$.
The terms in \eqref{W:susy} are generic quadratic terms containing
$\phi_a^\dag\phi_b$, $a\neq b$.
We will refer to this choice as \textit{quadratic} $W$.
It is the case of some supersymmetric models\,\cite{rasin.98}.
For special potentials with 4, 5 and 6 Higgs doublets, it was shown
that SCPV is not possible\,\cite{rasin.98}.
We are interested in generalizing this result to an arbitrary number of doublets.

First, we compute the derivatives for \eqref{W:susy}:
\eq{
\dd{W}{K_{ba}}=\la{ab}\,.
}
Then, Eq.\,\eqref{dV:K} yields
\eq{
\label{susy:K}
K_{aa}\dd{V_1}{K_{aa}}+\sum_{b}\la{ab}K_{ba}=0\,,
\quad a=1,\ldots,N\,.
}
Its imaginary part gives
\eq{
\label{susy:K:im}
\sum_{b}\la{ab}\im(K_{ba})=0\,,
\quad a=1,\ldots,N\,.
}

Instead of considering the real and imaginary parts independently, we can rewrite
\eqref{susy:K} as
\eq{
\label{susy:Kv}
v_a^2\dd{V_1}{K_{aa}}+\sum_{b}\la{ab}e^{i\theta_{ba}}v_av_b=0\,,
\quad a=1,\ldots,N\,,
}
where we have used
\eq{
\label{K->v}
K_{ab}=v_av_be^{i\theta_{ab}}\,,\quad
v_a\equiv\sqrt{K_{aa}}\,,~\theta_{ab}\equiv\theta_a-\theta_b\,.
}

After some straightforward manipulation we can rewrite \eqref{susy:Kv} as
\eq{
\sum_be^{-i\theta_a}\Big(
v_av_b\dd{V_1}{K_{aa}}\delta_{ab}+\la{ab}v_av_b\Big)e^{i\theta_{b}}
=0\,,\quad a=1,\ldots,N\,.
}
In matricial notation the equation is equivalent to
\eq{
\label{susy:Be=0}
\bbB\mtrx{1\cr e^{i\theta_{21}}\cr\vdots \cr e^{i\theta_{N1}}}=
\mtrx{0\cr 0\cr\vdots \cr 0}\,,
}
where the $N\times N$ real symmetric matrix $\bbB$ is defined by
\eq{
\label{susy:def:B}
(\bbB)_{ab}=v_av_b\dd{V_1}{K_{aa}}\delta_{ab}+\la{ab}v_av_b\,.
}
Explicitly,
\eq{
\bbB=
  \mtrx{
  v^2_1\dd{V_1}{K_{11}} & \la{12}v_1v_2  & \cdots & \la{1,N}v_1v_N\cr
  \la{21}v_2v_1 & v^2_2\dd{V_1}{K_{22}}   & \cdots & \la{2,N}v_2v_N\cr
  \vdots &  & \ddots & \vdots\cr
  \la{N,1}v_Nv_1 & \la{N,2}v_Nv_2 &\cdots  & v^2_N\dd{V_1}{K_{NN}}
  }\,.
}
Equation \eqref{susy:Be=0} implies that $\bbB$ as a function of $\{v_a\}$ should be
a singular matrix.

A noteworthy consequence of \eqref{susy:Be=0} should be pointed out:
the $N$ equations define $N$ polygons in the complex plane.
Each polygon, corresponding to the $a$-th row of \eqref{susy:Be=0}, is
made up of $N$ sides $|(\bbB)_{a1}|,|(\bbB)_{a2}|,\ldots,|(\bbB)_{aN}|$,  with
(external) angles $\theta_{i,i-1}=\theta_{i1}-\theta_{i-1,1}$ between the $i$-th
side $|(\bbB)_{ai}|$ and the extension of the $|(\bbB)_{a,i-1}|$ side; if some of
$(\bbB)_{ai}$ is not positive, we need to consider the angle $\theta_{i1}+\pi$
instead of $\theta_{i1}$.
Thus we have a nontrivial solution (SCPV) if we can find a set of $N$ polygons with
the same angles $\theta_{i,i-1}$.
Notice that among of the $\ums{2}N(N-1)$ sides $(\bbB)_{ai}$, only $N$ of them can
vary independently due to the independent $\{v_a\}$.

\subsection{Quartic $W$}

Let us now choose
\eq{
\label{W:4}
W=\ums{2}\sum_{a,b}
\la{ab}(\phi_a^\dag\phi_b)^2=\ums{2}\sum_{a,b}\la{ab}K_{ba}^2\,,
}
where $\la{ba}=\la{ab}$, $a\neq b$, is real and $\la{aa}=0$.
We will refer to this choice as \textit{quartic} $W$.
We assume terms such as $|\phi_a^\dag\phi_b|^2=|K_{ab}|^2$ are equivalent to
$K_{aa}K_{bb}$ for neutral VEVs and we include them into $V_1$.
Notice we are excluding terms such as $\phi_1^\dag\phi_2\phi_1^\dag\phi_3$ or
$\phi_1^\dag\phi_2\phi_3^\dag\phi_4$.
This can be achieved by imposing a $(Z_2)^N$ symmetry on the potential.
In this case
\eq{
\dd{W}{K_{ba}}=\la{ab}K_{ba}\,.
}

Equation \eqref{dV:K} is now
\eq{
\label{W:2:extr}
K_{aa}\dd{V_1}{K_{aa}}+\sum_{b}\la{ab}K_{ba}^2=0\,,\quad
a=1,\ldots,N\,.
}
If we use \eqref{K->v}, we obtain
\eq{
v^2_a\Big(
\dd{V_1}{K_{aa}}+\sum_{b}\la{ab}v^2_be^{i2\theta_{ba}}\Big)=0\,.
}
After some manipulations we can read
\eq{
\label{W:4:Ke=0}
\sum_be^{-i2\theta_a}\Big(
v^2_a\dd{V_1}{K_{aa}}\delta_{ab}+\la{ab}v^2_av^2_b\Big)e^{i2\theta_b} =0
\,.
}
We can rewrite the last equation as
\eq{
\label{W:4:Be=0}
\bbB\mtrx{1\cr e^{i2\theta_{21}}\cr\vdots \cr e^{i2\theta_{N1}}}=
\mtrx{0\cr 0\cr\vdots \cr 0}\,,
}
where
\eq{
\label{def:W:4}
(\bbB)_{ab}=v^2_{a}\dd{V_1}{K_{aa}}\delta_{ab}+\la{ab}v^2_av^2_b\,.
}

In complete analogy with Eq.\,\eqref{susy:Be=0}, Eq.\,\eqref{W:4:Be=0}
defines $N$ polygons with the only difference that now the angles are
$2\theta_{a1}$ and the off-diagonal terms $\bbB_{ab}$ depend quadratically on
the VEVs.
The $N$ sides of the $a$-th polygon are
$|(\bbB)_{a1}|,|(\bbB)_{a2}|,\ldots,|(\bbB)_{aN}|$,  with angles
$2\theta_{i,i-1}$ between the $i$-th side $|(\bbB)_{ai}|$ and the extension of the
$|(\bbB)_{a,i-1}|$ side.
We may have a nontrivial solution (SCPV) if we can find a set of $N$
polygons with the same angles $2\theta_{a1}$ and sides defined by
$(\bbB)_{ai}$.

\subsection{Constraints from nontrivial angles}

Let us analyze the polygon equations \eqref{susy:Be=0}, and, equivalently,
\eqref{W:4:Be=0}.
The equations themselves imply $\bbB$ must be singular:
\eq{
\label{rank:N-1}
\rank\bbB\le N-1\,.
}
However, if at least one polygon angle is nontrivial ($\theta_{21}\neq 0,\pi$) the
singularity is such that
\eq{
\label{rank<=N-2}
\rank\bbB\le N-2\,.
}
In fact, one nontrivial angle implies at least another nontrivial angle.
The reason for \eqref{rank<=N-2} is that we can write the first column of $\bbB$ as
a linear combination of the $N-1$ other columns as consequence of the real part of
\eqref{susy:Be=0}. For the same reason, the imaginary part of \eqref{susy:Be=0} ---
which does not depend on the first column of $\bbB$ ---  allows us to write the
second
column of $\bbB$ in terms of the $N-2$ other columns to the right of the second
column.

The equality in Eq.\,\eqref{rank:N-1} can be recovered in the absence of SCPV
where all $e^{i\theta_{i1}}=\pm 1$, i.e., a degenerate polygon confined to the real
line. In this case, there is no relation among the rows of $\bbB$, except from
\eqref{rank:N-1}. Therefore, the $N$ equations corresponding to the $N$ rows of
$\bbB$ are independent and can be used to find the $N$ quantities $v_a$.
This fact summarizes the relation between nontrivial angles (CP phases) and
constraints on the sides (VEVs).
These relations constrain the moduli $v_a$ and ultimately determines them.
If there is not enough structure in the potential to allow nontrivial polygons, then
there would be no CP breaking extremum.
In particular, for $N=2$ with either only quadratic or quartic potential,
 a two-sided polygon always resides in the real line and then
SCPV is not possible.

For the simplest nontrivial polygon, i.e., a triangle, \eqref{rank<=N-2} is enough
to find the nontrivial solution.
To form a nontrivial triangle (finite area), we need $\rank\bbB=1$. The geometrical
counterpart of such a requirement is that triangles with the same external (or
internal) angles must be similar. Conversely, similar triangles imply $\rank\bbB=1$.
This requirement can be translated to the fact that any $2\times 2$ minor of the
$3\times 3$ matrix $\bbB$ is zero. If we take the relations coming from
the vanishing of the three minors that only depend on one diagonal entry, then we
get three equations that define $v_1,v_2,v_3$. Once we get a solution for the VEVs
(sides of the triangles) we easily find the solutions for the CP phases (angles of
the triangles). This solution is unique when it exists.

Polygons with the same external angles are no longer guaranteed to be
similar for quadrilaterals or polygons with more sides.
The converse, however, remains true: if $N$ similar and noncontractible $N$-gons are
solutions for \eqref{susy:Be=0}, then
\eq{
\label{rank=1}
\rank\bbB=1\,.
}
This case is only possible if all $\la{ab}$ have the same sign.
The latter is just a necessary condition and, possibly for $N>3$, a solution
satisfying \eqref{rank=1} would require a special choice for the parameters of the
potential; see Sec.\,\ref{sec:B:finetune}.

\subsection{Algebraic consequences}
\label{sec:B:consequences}

We have seen in \eqref{rank<=N-2} that nontrivial CP phases require
$\rank\bbB\le N-2$.
Let us analyze its consequences.

Let us assume $\rank\bbB= N-2$. That means that $N-2$ rows of $\bbB$ are linearly
independent. Let them be the first $N-2$ rows. Since $\bbB$ is symmetric, we can
take the following set as the $(N-2)(N+3)/2$ independent elements of $\bbB$:
\eq{
\label{Bai:indep}
\bbB_{ai}\,,\quad a=1,\ldots,N-2,~i=1,\ldots,N\,~(i\ge a)\,.
}
The elements $\bbB_{ab}$, with $a,b> N-2$, can be written in terms of
\eqref{Bai:indep} as\,\cite{orbit.1}
\eq{
\label{Bab:dep}
\bbB_{ba}=\chi_b^\tp\big(\bbB^{(N-2)}\big)^{-1}\chi_a\,,
}
where $\bbB^{(N-2)}$ is the upper-left $(N-2)\times (N-2)$ submatrix of $\bbB$
and
\eq{\label{chidef}
\chi_a=(\bbB_{1a},\bbB_{2a},\cdots,\bbB_{N-2,a})^\tp\,,\quad
a,b=N-1,N\,.
}
Equations \eqref{Bab:dep} and \eqref{chidef} are the key equations in our discussion
of whether a multi-Higgs supersymmetric theory can support spontaneous CP violation
or not.
If $\rank\bbB<N-2$, more elements can be written in terms of a smaller set of
independent elements.
Equations \eqref{Bai:indep} and \eqref{Bab:dep} change accordingly.
The extreme case of $\rank\bbB=1$ is trivially satisfied by \eqref{Bab:dep}.

\subsection{Avoiding fine-tuning}
\label{sec:B:finetune}

Let us show here some situations that could lead to a fine-tuning of the parameters
of the potential unless they are imposed as a consequence of symmetries.
For that, we should notice that the off-diagonal elements of $\bbB$ depend on the
VEVs either as $\la{ab}v_av_b$ (quadratic $W$) or $\la{ab}v^2_av^2_b$ (quartic $W$).
The diagonal elements of $\bbB$ depends on the VEVs in more complex ways.
Let us assume quadratic $W$ to be specific; we only need to replace
$v_a\to v^2_a$ in the terms that contain $\la{ab}$.

We know that for triangles ($N=3$) Eq.\,\eqref{rank<=N-2} implies there is only
one possible solution for Eq.\,\eqref{Bab:dep}.
Let us think in the case of the next simplest case: $N=4$.
To have polygons with nontrivial angles, we need $\rank\bbB$ to be 1 or 2.
If $\rank\bbB=1$ --- the case of similar quadrilaterals --- all rows of $\bbB$ are
proportional to the first row.
Then any minor of order two ($2\times 2$ submatrix) is null.
In particular, we can find a $2\times 2$ submatrix above the diagonal that
does not contain elements from the diagonal of $\bbB$ itself.
Such a minor is
\eq{
\label{4:rank1}
\det\mtrx{\bbB_{13} & \bbB_{14}\cr \bbB_{23} & \bbB_{24}}=
\det\mtrx{\la{13} & \la{14}\cr \la{23} & \la{24}}v_1v_2v_3v_4=0\,.
}
Therefore, if we require that all the VEVs $v_a$ are nonzero, we are forced
to have
\eq{
\label{4:finetune}
\det\mtrx{\la{13} & \la{23}\cr \la{14} & \la{24}}=0\,.
}
Such a relation can be satisfied only by fine-tuning the parameters unless
 one finds some symmetry which guarantee this.
Barring such possibilities,  one can conclude that the only possibility for
SCPV to occur for the case of four Higgs doublets is to
have $\rank\bbB=2$; it is the maximal value where SCPV is possible.
We still have the possibility to set some VEVs to zero. But that means that the
problem reduces effectively to considering an $3\times 3$ submatrix of $\bbB$,
i.e., triangles; see Sec.\,\ref{sec:v=0}.

We can easily extend \eqref{4:rank1} to an arbitrary number $N$ of
doublets.
Many submatrices of the type of \eqref{4:rank1} can be constructed by
drawing a square inside the upper-right triangle of off-diagonal terms. We can see
that the largest of such submatrices has size $n\times n$ for $N=2n$ or $2n+1$.
Therefore,
\eq{
\label{no.finetune}
\text{no fine-tuning and nonzero VEVs}\implies
\rank\bbB\ge n\,, ~\text{for $N=2n$ or $N=2n+1$}\,.
}
This condition, together with \eqref{rank<=N-2}, reduces the number of cases we have
to treat.
In particular, for $N>3$, similar polygons can not be solutions to the
extremization problem \eqref{susy:Be=0} [and \eqref{W:4:Be=0}] if we avoid
fine-tuning or null VEVs.

\subsection{Null VEVs}
\label{sec:v=0}

Let us treat the case of null VEVs.
Suppose $v_N=0$. Then the matrix $\bbB$ reduces from a $N\times N$ matrix to an
effective $(N-1)\times(N-1)$ matrix $\bbB'$ because the $N$-th row and the
$N$-th column of $\bbB$ is now entirely null.
If the rest of the VEVs $v_a$ are nonzero, we can use the argument used to obtain
the constraint \eqref{rank<=N-2} and apply it to the effective non-null submatrix
$\bbB'$.
We conclude that
\eq{
\label{rank:vN=0}
\rank\bbB'\le (N-1)-2\,;
}
the rank of $\bbB'$ is reduced by at least two units and the same conditions of
Secs.\,\ref{sec:B:consequences} and \ref{sec:B:finetune} apply as if the relevant
quantity were $\bbB'$.
If a number $p$ of VEVs are null, we can extract the effective submatrix $\bbB'$ of
size $m\times m$, $m=N-p$, by eliminating the $p$ rows and columns made entirely of
zeros.
If we require at least one nontrivial relative angle (corresponding to two nonzero
VEVs), we still need
\eq{
\label{rank:va=0}
\rank\bbB=\rank\bbB'\le m-2\,.
}

The condition \eqref{no.finetune} also changes to
\eq{
\label{no.finetune:2}
\text{no fine-tuning}\implies
\rank\bbB\ge n\,, ~\text{for $m=2n$ or $m=2n+1$}\,.
}

\section{Applications to NHMSSM}
\label{sec:susy:gen}

Let us show here that there can be no SCPV at tree level in N-Higgs-doublets
supersymmetric extensions of the SM if (i) we avoid fine-tuning of the parameters
of the potential and (ii) the quartic terms come solely from D-terms.
The proof is essentially an application of the constraints explained in
Sec.\,\ref{sec:B:consequences}.

We consider two cases: (A) an arbitrary number of $N=2n$ Higgs doublets and (B)
$n_f=3$ effective Higgs doublets coming from the sleptons $\tilde{L}_i$,
$i=1,2,3$, in addition to an arbitrary number of $2n$ Higgs doublets; the effective
number of doublets is $N=2n+n_f$.
The latter case is usually referred to as R-parity violating
models\,\cite{rasin.98}.

In both types of models, we label the $n+n_f$ ($n_f=0$ or $n_f=3$) doublets with
hypercharge $Y=-1$ with indices $a=1,\ldots,n+n_f$, and the $n$ doublets
with hypercharge $Y=1$ with the indices $a=n+n_f+1,\ldots,N$; $N=2n+n_f$.
Moreover, for $Y=1$ doublets we just denote $H_{a}\to \phi_a$ while for $Y=-1$
doublets we associate $H_a\to \tilde{\phi}_a=i\sigma_2\phi_a^*$.
All the effective doublets now have hypercharge $Y=1$ and they are all denoted
by $\{\phi_a\}$, $a=1,\ldots,N$; $N=2n+n_f$.

\subsection{$N=2n$ Higgs doublets}
\label{sec:susy:gen:2n}

Let us consider $N=2n$ Higgs doublets. We are assuming that the quartic part of the
potential comes only from the D-term
\eq{
\label{susy:gen:D:2n}
V_D=\ums{2}G\big[|\phi_1|^2+|\phi_2|^2+\cdots+|\phi_{n}|^2-|\phi_{n+1}|^2-\cdots
-|\phi_{2n}|^2\big]^2\,,
}
where $G=(g^2+g^{\prime 2})/4$.
We also assume the VEVs of the doublets do not break electric charge.
The potential \eqref{susy:gen:D:2n} is translated, in the notation of \eqref{V:2|4},
into
\eq{
\label{cab:susy:gen}
(c_{ab})=G\mtrx{A_n & -A_n \cr -A_n & A_n}\,,
}
where $A_n=\ones{n}{n}$ is an $n\times n$ matrix with $(A_n)_{ab}=1$ for all
entries.
The general potential will have the form \eqref{V:2|4} with quadratic $W$ as
\eqref{W:susy}. Except for $(c_{ab})$, which is defined in \eqref{cab:susy:gen},
all parameters of the potential are generic as long as they can provide a neutral
vacuum and a potential bounded from below.

One peculiarity of real potentials with quadratic $W$ is that we can rewrite
$\bbB$ in Eq.\,\eqref{susy:def:B} as
\eq{
\label{W:2:B->tB}
\bbB=\diag(v_1,v_2,\ldots,v_N)\,\tB\,\diag(v_1,v_2,\ldots,v_N)\,,
}
where
\eq{
\label{W:2:def:tB}
\tB_{ab}= \delta_{ab}(\mu_a+\sum_d c_{ad}v^2_d)+\la{ab}\,.
}
The matrix $\tB$ depends on the VEVs $v_a$ only in the diagonal
terms and such a dependence occurs only through the terms $\sum_d
c_{ad}v^2_d$. In our particular case, the matrix $(c_{ab})$
in Eq.\,\eqref{cab:susy:gen} is so restrictive that it allows us to write
\eq{
\sum_d c_{ad}v^2_d=\eps_af(v^2)\,,
}
where $\eps_a=1$ for $1\le a\le n$ and $\eps_a=-1$ for $n<a\le N$.
The quantity $f(v^2)$ is the combination
\eq{
\label{susy:gen:def:f}
f(v^2)\equiv G\sum_{a=1}^n(v^2_a-v^2_{n+a})\,.
}
We will see that the dependence of $\tB$ only on the combination $f(v^2)$ is too
restrictive to allow SCPV, unless a fine-tuning of the parameters is allowed.

We first allow the possibility that some number $p\ge 0$ of VEVs are null, and no
more.
Let us eliminate the null rows and columns and extract the effective submatrix
$\bbB'$ of size $m\times m$, where $m=N-p$.
The requirement of at least one nontrivial CP phase imposes \eqref{rank:va=0}, i.e.,
\eq{
\label{susy:gen:def:r}
r\equiv\rank\bbB=\rank\bbB'\le m-2\,.
}
The possibility that $\bbB'$ has more null rows (columns) is excluded because that
would require either more null VEVs or entire rows of $(\la{ab})$ to be
null.

Let us rearrange the rows (columns) of $\bbB$ in such a way that the first $m$
rows (columns) correspond to the rows (columns) of $\bbB'$.
We also rearrange the rows (columns) of $\bbB'$ in such a way that the first $m-2$
rows (columns) are linearly independent.
We relabel $\mu_a,c_{ab},\la{ab}$ accordingly.
All this rearrangement maintains the diagonal entries of $\bbB$ in the diagonal
and we still have
\eq{
\bbB_{aa}=v^2_a\big[\mu_a+\eps_a f(v^2)\big]\,,\quad
a=1,\ldots, m\,,
}
where $f(v^2)$ is given in \eqref{susy:gen:def:f} and $\eps_a=\pm 1$.

Let us suppose now that $m\ge 3$. Nontrivial CP phases are not possible for
$m=1$ or $m=2$, even for generic potentials with quadratic or quartic $W$.
Take $\bbB^{(r)}$ as the $r\times r$
upper-left submatrix of $\bbB$ (or $\bbB'$); $r\le m-2$ by \eqref{susy:gen:def:r}.
The matrix $\bbB^{(r)}$ is nonsingular by construction.
Then a generalization of Eq.\,\eqref{Bab:dep} applies at least to $\bbB_{mm}$,
$\bbB_{m-1,m-1}$ and $\bbB_{m-1,m}$, i.e.,
\eqali{
\label{susy:gen:Bmm}
\bbB_{m-1,m-1}&=
\chi_{m-1}^\tp(\bbB^{(r)})^{-1}\chi_{m-1}\,,\cr
\bbB_{m,m}&=
\chi_{m}^\tp(\bbB^{(r)})^{-1}\chi_{m}\,,\cr
\bbB_{m-1,m}&=
\chi_{m-1}^\tp(\bbB^{(r)})^{-1}\chi_{m}\,,
}
where
\eq{
\chi_a=(\bbB_{1a},\bbB_{2a},\ldots,\bbB_{r,a})^\tp\,,\quad a=m-1\text{ or }m\,.
}

We can also define $\tchi_a$ as
\eq{
\tchi_a\equiv(\la{1a},\la{2a},\ldots,\la{r,a})^\tp\,,
}
so that
\eq{
\chi_a=v_a\diag(v_1,v_2,\ldots,v_{r})\tchi_a\,,\quad a=m-1,m\,.
}
Then Eq.\eqref{susy:gen:Bmm} can be rewritten as
\eqali{
\label{susy:gen:BNN:2}
v^2_{m-1}\tB_{m-1,m-1}&=v^2_{m-1}\,
\tchi_{m-1}^\tp\big(\tB^{(r)}\big)^{-1}\tchi_{m-1}\,, \cr
v^2_m\tB_{m,m}&=v^2_m\,
\tchi_{m}^\tp\big(\tB^{(r)}\big)^{-1}\tchi_{m}\,,\cr
v_{m-1}v_m\la{m-1,m}&=v_{m-1}v_m\,
\tchi_{m-1}^\tp\big(\tB^{(r)}\big)^{-1}\tchi_{m}\,,
}
where
\eq{
\tB^{(r)}=
  \mtrx{
  \mu_1+\eps_1 f(v^2) & \la{12} &  \cdots & \la{1r}\cr
  \la{21} & \mu_1+\eps_2 f(v^2) &  \cdots & \la{2r}\cr
  \vdots & \vdots &\ddots & \vdots \cr
  \la{r,1} & \la{r,2} &\cdots & \mu_{r}+\eps_r f(v^2)
  }\,.
}

Now we use the fact that $v_m$ and $v_{m-1}$ are non-null by hypothesis. We then
obtain
\subeqali{susy:gen:BNN:3}{
\label{susy:gen:B:N-1}
\mu_{m-1}+\eps_{m-1}f(v^2)&=
\tchi_{m-1}^\tp\big(\tB^{(r)}\big)^{-1}\tchi_{m-1}\,, \\
\label{susy:gen:B:N}
\mu_{m}+\eps_m f(v^2)&=
\tchi_{m}^\tp\big(\tB^{(r)}\big)^{-1}\tchi_{m}\,\\
\label{susy:gen:B:mm-1}
\la{m-1,m}&=
\tchi_{m-1}^\tp\big(\tB^{(r)}\big)^{-1}\tchi_{m}\,.
}
Notice that these three equations only depend on the VEVs through the combination
in $f(v^2)$.
Therefore if we solve Eq.\eqref{susy:gen:B:N-1} for $f(v^2)$ in terms of the
parameters of the potential, Eqs.\,\eqref{susy:gen:B:N} and \eqref{susy:gen:B:mm-1}
will give us two relations between the parameters of the potential.
Such relations will be only satisfied by fine-tuning the parameters and is not
stable under radiative conditions. Therefore, we conclude that CP phases must vanish
and hence there is no spontaneous CP violation.

If $r<m-2$ strictly, we can add to the list of equations in \eqref{susy:gen:BNN:3}
$m-2-r$ more relations coming from Eq.\,\eqref{Bab:dep}
applied to the diagonal entries $\bbB_{aa}$, $r< a\le m-2$.
We can also add more relations associated to the off-diagonal entries $\bbB_{ab}$,
$r< a\le m-2$, $r< b\le m-2$, $a<b$.
Then more relations have to be satisfied simultaneously by the only combination in
$f(v^2)$ and the fine-tuning is worsened.

Therefore, we conclude that the quartic terms --- coming from D-terms --- in
supersymmetric multi-Higgs-doublet extensions of the SM do not have enough structure
to break CP spontaneously, independently of the number of doublets $N=2n$.

\subsection{R-parity-violating models}
\label{susy:gen:R}

Let us consider $N=2n+n_f$ doublets, $n_f=3$ coming from the sleptons $\tilde{L}_i$,
one for each family $i$; see Ref.\,\cite{rasin.98}.

We can still assume that the quartic terms, for neutral VEVs, comes solely from the
D-term
\eq{
\label{susy:gen:D:Li}
V_D=\ums{2}G\Big[\sum_{a=1}^{n+n_f}|\phi_a|^2
    -\sum_{a=n+n_f+1}^{N}|\phi_{a}|^2\Big]^2\,.
}
The matrix $(c_{ab})$ is now
\eq{
\label{cab:susy:gen:R}
(c_{ab})=G\mtrx{A_{n+n_f} & -\ones{(n+n_f)}{n}\cr
          -\ones{n}{(n+n_f)} & A_{n}}\,.
}
As in \eqref{cab:susy:gen}, the matrices $\ones{n}{m}$ are $n\times m$ matrices
where all entries are unity.

Now, the diagonal entries of $\tB$ in \eqref{W:2:def:tB} still depend only on
one combination of the VEVs,
\eq{
\sum_d c_{ad}v^2_d=\eps_af(v^2)\,,
}
where now
\eq{
\label{susy:gen:f:R}
f(v^2)\equiv G\Big[\sum_{a=1}^{n+n_f} v^2_a-\sum_{a=n+n_f+1}^{N}v^2_{a}\Big]\,.
}
The factors $\eps_a$ are $\eps_a=+1$ for $1\le a\le n+n_f$ and $\eps_a=-1$ for
$n+n_f<a\le N$.
Therefore the arguments used in Eqs.\,\eqref{susy:gen:def:r}--\eqref{susy:gen:B:mm-1}
still apply and it is not possible to find solutions to the extremum equations with
nontrivial CP phases.
Thus it is not possible to have SCPV in supersymmetric multi-Higgs extensions of
the SM with spontaneous R-parity violation.

\subsection{4-HMSSM}
\label{sec:4HMSSM}

In this subsection, we illustrate the application of our technique to the case of
4-HMSSM
(i.e., 4-Higgs-doublet supersymmetric extension of the SM)\,\cite{rasin.98}.
In this case, the quartic part of the potential \eqref{susy:gen:D:2n} is simply
\eq{
V_D=\ums{2}G\big[|\phi_1|^2+|\phi_2|^2-|\phi_3|^2-|\phi_4|^2\big]^2\,.
}
This is equivalent to considering the matrix $(c_{ab})$ as
\eq{
\label{susy:4:c}
(c_{ab})=G\mtrx{
    1 &1 &-1 &-1\cr
    1 &1 &-1 &-1\cr
    -1 &-1 &1 &1\cr
    -1 &-1 &1 &1\cr
    }\,,
}
where $G\equiv (g_1^2+g_2^2)/4$.
The function $f$ \eqref{susy:gen:def:f} in this case is
\eq{
f(v^2)\equiv G[v^2_1+v^2_2-v^2_3-v^2_4]\,.
}

Allowing for null VEVs, we need at least three non-null VEVs to have at
least two nontrivial CP phases. Therefore, the effective non-null matrix $\bbB'$
would have size $m=3$ or $m=4$.
For both cases, nontrivial CP phases require $r=\rank\bbB\le m-2$.
That in turn, implies the relations \eqref{susy:gen:BNN:3}.

For $m=4$ (no null VEVs) and $\rank\bbB=2$, considering the first two rows of $\bbB$
to be linearly independent, we have explicitly
\eqali{
\label{4hdm:finetune:42}
\mu_3-f(v^2)=&
\mtrx{\la{13}&\la{23}}
    \mtrx{\mu_1+f(v^2) & \la{12} \cr
      \la{21} &\mu_2+f(v^2)\cr
    }^{-1}
\mtrx{\la{13}\cr\la{23}}\,,\cr
\mu_4-f(v^2)=&
\mtrx{\la{14}&\la{24}}
    \mtrx{\mu_1+f(v^2) & \la{12} \cr
      \la{21} &\mu_2+f(v^2)\cr
    }^{-1}
\mtrx{\la{14}\cr\la{24}}\,,\cr
\la{34}=&
\mtrx{\la{13}&\la{23}}
    \mtrx{\mu_1+f(v^2) & \la{12} \cr
      \la{21} &\mu_2+f(v^2)\cr
    }^{-1}
\mtrx{\la{14}\cr\la{24}}\,.
}

For $m=4$ and $\rank\bbB=1$, \eqref{4hdm:finetune:42} is replaced by
\subeqali{4hdm:finetune:41}{
\label{4hdm:finetune:41:2}
\mu_2+f(v^2)=&\frac{\la{12}^2}{\mu_1+f(v^2)}\,,\\
\label{4hdm:finetune:41:3}
\mu_3-f(v^2)=&\frac{\la{13}^2}{\mu_1+f(v^2)}\,,\\
\mu_4-f(v^2)=&\frac{\la{14}^2}{\mu_1+f(v^2)}\,,
}
and
\subeqali{4hdm:finetune:41:off}{
\label{4hdm:finetune:41:23}
\la{23}&=\frac{\la{12}\la{13}}{\mu_1+f(v^2)}\,,\\
\la{24}&=\frac{\la{12}\la{14}}{\mu_1+f(v^2)}\,,\\
\la{34}&=\frac{\la{13}\la{14}}{\mu_1+f(v^2)}\,.
}

For $m=3$ and $\rank\bbB=1$, assuming $v_4=0$, the relations for $\bbB_{22}$ and
$\bbB_{33}$ are the same as Eqs.\,\eqref{4hdm:finetune:41:2} and
\eqref{4hdm:finetune:41:3}; the relation for $\bbB_{23}$ is the same as
\eqref{4hdm:finetune:41:23}.
If other VEVs are null instead of $v_4$, we obtain similar relations.

In all cases,  fine-tuning is necessary to satisfy the requirements of
nontrivial CP phases.

\section{Conclusions}
\label{sec:conclusions}

To summarize, we have seen that the possibility of SCPV in NHDM potentials where complex
combinations $\phi^\dag_a\phi_b$ appear only on either quadratic
($\phi^\dag_a\phi_b$) or quartic $(\phi^\dag_a\phi_b)^2$ parts
can be related to the existence of $N$ nontrivial polygons defined by a $N\times N$
real matrix $\bbB$ which is a function of the moduli $|\phi_a|=v_a$ as in
\eqref{susy:def:B} or \eqref{def:W:4}. These polygons generalize the triangle
described in \cite{branco} for the Weinberg 3-Higgs-doublet model.
For models with a moderate number of doublets, the characterization of the
possible extrema through polygons should be still useful to guide the
search for solutions.

As a direct application of these ideas, we have shown that in
arbitrary Higgs-doublets extension of MSSM, it is impossible to break CP
spontaneously at tree level unless we allow fine-tuned relations
between tree level parameters of the theory. This result remains
true even if the sneutrinos acquire VEVs so that R-parity is
spontaneously broken. This generalizes the earlier result in
\cite{rasin.98} for two or three pairs of Higgs doublets to the
case with an arbitrary number of them. A simple corollary of this
result is that in supersymmetric left-right models with arbitrary number of
bidoublets  where $\rm B-L$ is broken by $\rm B-L=2$ triplets,
there is no spontaneous CP violation in the bidoublet sector. This
result has application to a solution of the strong CP problem
without the need for an axion (see, for instance,\,\cite{MR}).

\acknowledgements

The work of C.C.N. was partially supported by Brazilian
{\em Conselho Nacional de Desenvolvimento Científico e Tecnológico} (CNPq)
and {\em Fundação de Amparo à Pesquisa do Estado de São Paulo} (Fapesp).
C.C.N. also thanks the \emph{Maryland Center for Fundamental Physics} for the
hospitality during the development of this work. The work of R.N.M. was supported
by the NSF grant PHY-0968854.


\end{document}